\documentclass[%
 reprint,
 amsmath,amssymb,
 aps,
]{revtex4-2}

\usepackage{graphicx}% Include figure files
\usepackage{dcolumn}% Align table columns on decimal point
\usepackage{bm}% bold math
\usepackage{xcolor}
\begin{document}

\preprint{APS/123-QED}

\title{Dark Energy in the $w$–$c_s^2$ Plane}

\author{Oem Trivedi, Robert J. Scherrer, Alfredo Gurrola}
\affiliation{Department of Physics and Astronomy, Vanderbilt University, Nashville, TN, 37235, USA}
\email{Email: oem.trivedi@vanderbilt.edu \\ Email : robert.scherrer@vanderbilt.edu \\ Email: alfredo.gurrola@vanderbilt.edu }

\date{\today}% It is always \today, today,
             %  but any date may be explicitly specified

\begin{abstract}
We introduce a unified framework for dark energy diagnostics based on the joint phase space of the equation of state $w$ and the sound speed $c_s^2$. This $w$-$c_s^2$ plane provides a minimal extension beyond background cosmology, capturing both the expansion and perturbative properties within a single representation. Building on this, we then define the microphysical flow parameter $\mathcal{F} = dc_s^2/dw$, which encodes the dynamical relation between background evolution and perturbative response.  We derive a direct connection between $\mathcal{F}_0$ (the present-day value of $\mathcal{F}$), $H_0$, and $\sigma_8$. We show that the microphysical flow parameter enables a hierarchy of increasingly stringent consistency tests that significantly reduce the viable dark energy theory space. We further demonstrate how trajectories in the $w$-$c_s^2$ plane distinguish models that are nearly degenerate at the level of $w(a)$, including canonical quintessence, Chaplygin gas models, and noncanonical scalar field realizations. This provides a compact phenomenological bridge between dark energy microphysics and future perturbation-sensitive observations, making $\mathcal{F}_0$ a useful discriminator of the kinetic structure which is underlying cosmic acceleration.
\end{abstract}

\maketitle

\section{Introduction}
It is fair to say that contemporary cosmology is undergoing one of the most dynamic periods in its history. Both theoretical developments and observational breakthroughs have advanced to a point where the Universe can now be studied across vast dynamical scales with remarkable accuracy. In particular, the breadth of possible dark energy models \cite{de1SupernovaSearchTeam:1998fmf,de2Li:2012dt,de3Li:2011sd,de4Mortonson:2013zfa,de5Frusciante:2019xia,de6Huterer:2017buf,de7Vagnozzi:2021quy,de8Adil:2023ara,de9Feleppa:2025clx,de10DiValentino:2020evt,de11Nojiri:2010wj,de12Nojiri:2006ri,de13Trivedi:2023zlf,de14Trivedi:2022svr,de15Trivedi:2024inb,de16Trivedi:2024dju,re2Bamba:2012cp,re3Andriot:2025los} has significantly broadened the theoretical landscape of late-time cosmic evolution, while persistent anomalies such as the $H_0$ and $\sigma_8$ tensions \cite{ht1DiValentino:2021izs,ht2Clifton:2024mdy,s81kazantzidis2018evolution,s82amon2022non,s83poulin2023sigma,s84Ferreira:2025lrd,hubtenCai:2026swf} continue to expose potential cracks in the standard $\Lambda$CDM paradigm. At the same time, cosmology is increasingly becoming a data-rich science. Upcoming and ongoing observational efforts including next-generation CMB experiments, large scale structure surveys, high redshift probes, gravitational wave observatories, and precision mapping of the cosmic web are expected to place stringent constraints on both early and late universe physics \cite{t1jha2019next,t2Shanks:2015lda,t3chandler2025nsf,t4Euclid:2024yrr,t5WST:2024rai,t6CosmoVerseNetwork:2025alb,t7COMPACT:2022gbl}. The interplay between this expanding observational capability and a rapidly evolving theoretical framework is driving the field toward deeper questions regarding the fundamental nature of spacetime, gravity, and cosmic acceleration.

The dark energy equation of state parameter, $w=p_{\rm DE}/\rho_{\rm DE}$, provides the most commonly used phenomenological characterization of cosmic acceleration and determines the evolution of the homogeneous dark energy background. Observational efforts have hence primarily focused on constraining $w$, particularly its present value and possible redshift dependence \cite{Planck:2018vyg,desi1lodha2025extended,desi2abdul2025desi,desi3AtacamaCosmologyTelescope:2025blo,desi4DESI:2025zpo,desi5DESI:2023dwi}. 
This has led to the development of numerous graphical and diagnostic approaches aimed at extracting physical insight from observational data, including the widely used $w$-$w'$ phase space \cite{ww1Caldwell:2005tm,ww2Scherrer:2005je,ww3Janssen:2009nz,ww4Linder:2006xb,ww5Malekjani:2012tn}, the $\mathrm{Om}(z)$ diagnostic \cite{om1Sahni:2008xx,om2Zunckel:2008ti,om3Lu:2008hp,om4Thakur:2021mot}, and the statefinder hierarchy \cite{sta1Sahni:2002fz,sta2Alam:2003sc,sta3Zhang:2005yz,sta4Gao:2010ia,sta5Sharif:2012uvx,sta6Carrasco:2023imi}. These methods have proven to be powerful in distinguishing between broad classes of models, such as the standard cosmological constant with cold dark matter ($\Lambda$CDM), quintessence, and certain modified gravity scenarios, particularly by focusing on deviations from $\Lambda$CDM in the expansion history.

However, despite their utility, these approaches share a common structural limitation, namely that they are fundamentally rooted in the background cosmology. The $w$-$w'$ plane, for instance, provides a useful classification of dynamical dark energy models by examining the evolution of $w$ with respect to the logarithmic scale factor, thereby distinguishing between thawing and freezing behaviors. Similarly, the $\mathrm{Om}(z)$ diagnostic offers a geometrical test of $\Lambda$CDM by comparing the Hubble parameter at different redshifts, while the statefinder parameters $(r,s)$ and their extensions probe higher derivatives of the scale factor to characterize deviations from simple models. All of these diagnostics are ultimately constructed from the expansion history ($H(z)$ and its derivatives) and therefore inherit the same intrinsic limitation: they are all insensitive to the perturbative properties of dark energy.

Even a precise determination of $w$ does not fully specify the physical nature of dark energy, since models with nearly identical background expansion histories can exhibit substantially different perturbative behavior. A complementary quantity is the squared sound speed, $c_s^2$, which governs the propagation and clustering of dark energy perturbations and thereby probes its underlying kinetic and microphysical structure. Despite its central importance, obtaining observational constraints on the sound speed of dark energy has proven to be highly nontrivial. Current limits are largely derived from late-time, integrated observables such as the Integrated Sachs–Wolfe (ISW) effect \cite{csc1Hannestad:2005ak,csc2de2010measuring,csc3ballesteros2010dark,csc4linton2018variable,csc5bean2004probing,csc6eisenstein2005dark,csc7sergijenko2015sound,ss1yang2025clustering,spc1Erickson:2001bq,spc2Weller:2003hw,spc3Xia:2007km,spc4Ballesteros:2010ks,spc5basse2012confronting,stef1Perkovic:2024vgp,stef2Perkovic:2022ykp,stef3Perkovic:2020eju,stef4Perkovic:2020mph,stef5Perkovic:2019vxm,stef6Perkovic:2018pzv,stef7Caplar:2012ed}. These observables depend on line of sight integrals over evolving gravitational potentials and therefore provide only indirect sensitivity to the underlying perturbation dynamics. As a result, the regions of parameter space associated with the kinetic sector and the propagation of scalar modes remain only weakly constrained.  We recently introduced a qualitatively different observational strategy aimed at directly probing the microphysics of cosmic acceleration \cite{soundspeed1,soundspeed2}.  In particular, we showed that a lunar based interferometric setup such as the Lunar Interferometer for Laser Astronomy (LILA) \cite{lilajani2025laser} can access ultra-low frequency gravitational signals corresponding to horizon-scale scalar potentials. This would provide, in principle, a direct measurement of the dark energy sound speed through real-time observations of the evolution of large-scale gravitational potentials, opening a new observational window into the perturbative sector of dark energy.

In light of these developments, it is clear that a complete and physically meaningful characterization of dark energy requires a framework that goes beyond single-perspective descriptions such as $w(z)$. The existing suite of diagnostic plots, while valuable, effectively provides a one dimensional view of a fundamentally multidimensional problem. What is required is a more comprehensive approach that simultaneously captures both the background evolution and the perturbative behavior of dark energy, allowing for a unified and discriminating classification of models.  Hence, we introduce here the $w - c_s^2$ plane and a big motive behind that lies in a simple fact. The EOS parameter $w$ controls encodes information about background expansion with dark energy and while the sound speed $c_s^2$ controls information about dark energy perturbations, and hence, a paired set of them would represent the lowest-order complete phenomenological description of dark energy at linear order. 
Our goal is not to replace existing methods, but to extend them into a framework that reflects the full physical content of the dark energy sector. This motivates the search for new graphical representations that incorporate multiple complementary aspects of dark energy dynamics and that can fully exploit the capabilities of upcoming observational probes. Beyond introducing the $w-c_s^2$ plane, we also define a new cosmological parameter which we call the "Microphysical Flow Parameter" denoted by $\mathcal{F}$, that characterizes how the perturbative sector evolves relative to the background. We will also demonstrate how $\mathcal{F}$ provides powerful consistency tests and connects naturally future observations that are sensitive to perturbative effects. We would also show how familiar dark energy scenarios populate the $w-c_s^2$ plane and how they can be better bifurcated in detail using the plane and the flow parameter. 

\section{The $w$-$c_s^2$ Plane}
\subsection{Defining the Plane}
In light of the previous discussion, it is evident that any meaningful attempt to characterize dark energy must go beyond purely background-based diagnostics and incorporate information about its perturbative behavior. A natural and minimal extension of existing approaches is to consider a two-dimensional phase space defined by the equation of state parameter, $w(z)$, and the effective sound speed, $c_s^2(z)$.  In terms of the dark energy pressure $p_{DE}$ and density $\rho_{DE}$, these are given by
\begin{equation}
\label{wdef}
w = \frac{p_{DE}}{\rho_{DE}},
\end{equation}
and
\begin{equation}
\label{csdef}
c_s^2 = \frac{dp_{DE}}{d\rho_{DE}}.
\end{equation}
These two quantities together provide a complete description of dark energy at linear order as $w(z)$ determines the homogeneous evolution of the energy density and the expansion history, while $c_s^2(z)$ governs the response of pressure perturbations to density fluctuations and hence controls the clustering properties of the dark energy component.

This choice is not arbitrary but is both theoretically well motivated and observationally timely. From a theoretical point of view, the pair $(w, c_s^2)$ contains the essential degrees of freedom of a general fluid or effective field theory description of dark energy.  Different models can be degenerate in $w(z)$ at the background level, but once perturbations are included, $c_s^2$ provides an additional axis along which these models can be distinguished. From an observational perspective, upcoming surveys and novel probes such as ultra low frequency interferometers like LILA are expected to provide direct sensitivity to the dynamics of large scale gravitational potentials, opening the possibility of constraining $c_s^2(z)$ independently of $w(z)$. This makes the $w$-$c_s^2$ plane a natural arena in which to interpret future data and to classify dark energy models in a unified manner. In the context of dark matter physics, models on generalized dark matter have emphasized the need to jointly understand the evolution of the EOS parameter and sound speed  \cite{gdm1Hu:1998kj,gdm2Hu:2000ke,gdm3Mocz:2023adf,gdm4Hu:2007pj}. However, a proper classification of models has not been explored using such a plane in dark matter phenomenology too.

The cleanest way to organize this plane is by the signs of $1+w$ and $c_s^2$. The line $w=-1$ separates quintessence like behavior from phantom like behavior, while the line $c_s^2=0$ separates perturbatively stable propagation from gradient instability. The cosmological constant itself would lie on the boundary $w=-1$, with no propagating dark energy perturbation in the usual $\Lambda$CDM limit. We therefore divide the plane into four main regimes and then place the more specific model classes inside the appropriate regime. This division can be organized as follows : 

\begin{itemize}

\item \textbf{$w>-1,\; c_s^2>0$: nonphantom and perturbatively stable dark energy}
\\
This is the most conservative and theoretically well-behaved region of the plane. The condition $w>-1$ corresponds to nonphantom dark energy, so the null energy condition is not violated at the effective background level. The condition $c_s^2>0$ ensures that pressure gradients oppose the growth of perturbations rather than producing a gradient instability. If, in addition, $w<-1/3$, models in this quadrant can drive accelerated expansion while remaining perturbatively stable.

\begin{itemize}

\item \textit{$w\approx -1,\; c_s^2\approx 1$:} This is the smooth cosmological constant like limit, as models in this region mimic $\Lambda$CDM both in the expansion history and in the perturbative sector. A sound speed near unity gives efficient pressure support, suppressing dark energy perturbations on subhorizon scales. The dark energy component therefore remains effectively homogeneous and mainly affects the Universe through the background expansion.

\item \textit{$w\approx -1,\; c_s^2\ll 1$:} This is an especially interesting near-$\Lambda$CDM regime, as here the background expansion can be almost indistinguishable from a cosmological constant, but the perturbation sector can be very different. A small sound speed weakens pressure support and allows dark energy perturbations to survive on large scales and such models can affect the evolution of gravitational potentials, the integrated Sachs-Wolfe effect and potentially the ultra low frequency strain spectrum probed by future interferometric observatories, even if their distance-redshift behavior remains close to $\Lambda$CDM.

\item \textit{$w>-1,\; c_s^2\approx 1$:} This is the standard smooth quintessence like regime and it is typically realized by canonical scalar fields with $P(\phi,X)=X-V(\phi)$ for which $c_s^2=1$ identically. The equation of state can evolve in time and give a dynamical expansion history, but the perturbations are efficiently damped. These models are hence mostly tested through background observables and through their indirect effect on growth via the expansion rate.

\item \textit{$w>-1,\; c_s^2\ll 1$:} This corresponds to clustering quintessence or more general noncanonical dark energy, including $k$-essence and effective field theory realizations with suppressed sound speed. In this case the background sector is dynamical and the perturbative sector is also active. Such models can modify both the expansion history and the growth of structure, providing a way to distinguish them from smooth quintessence even when the histories of $w(z)$ are similar.

\item \textit{$w>-1$,\; $c_s^2>1$:} This does not define a separate quadrant, but it is a physically distinctive part of the positive sound speed half plane. Values larger than unity often point toward highly noncanonical behavior, higher derivative structure or an effective description whose causal interpretation must be handled carefully. An observational preference for this region would point strongly away from simple canonical scalar field models.

\end{itemize}

\item \textbf{$w<-1,\; c_s^2>0$: phantom like but perturbatively stable dark energy}

This quadrant corresponds to dark energy with phantom like background evolution but stable pressure perturbations. The condition $w<-1$ implies effective violation of the null energy condition and leads to an energy density that increases as the Universe expands. For a constant equation of state $\rho_{DE}\propto a^{-3(1+w)}$, so $w<-1$ produces growing dark energy density and can lead, in simple models, to a future Big Rip singularity. The condition $c_s^2>0$ means that the perturbations are not subject to a gradient instability, although the background dynamics are already exotic. We can further understand it in the following sub-cases:
\begin{itemize}

\item \textit{$w<-1,\; c_s^2\approx 1$:} This is the smooth phantom like regime and these models can resemble $\Lambda$CDM in their perturbative behavior because pressure support remains efficient, but their background evolution is qualitatively different. Since canonical scalar fields cannot realize $w<-1$ without pathologies, this region usually requires noncanonical scalar fields, multiple fields, modified gravity or an effective fluid description.

\item \textit{$w<-1,\; c_s^2\ll 1$:} This is the clustering phantom like regime and here both the background and perturbative sectors are nonstandard. The background undergoes super accelerated expansion, while the small sound speed permits dark energy perturbations to cluster on large scales. This combination can produce strong signatures in the evolution of gravitational potentials and in the growth history. Such models are likely to be highly constrained, but if realized, they would provide clear evidence for physics beyond simple scalar field dark energy.

\item \textit{$w<-1,\; c_s^2>1$:} This would combine phantom background evolution with highly noncanonical perturbative propagation. Such a regime is theoretically extreme, but it remains useful to keep it in the classification because effective theories or modified gravity models can sometimes produce unusual inferred sound speeds.

\end{itemize}

\item \textbf{$w<-1,\; c_s^2<0$: phantom like and perturbatively unstable dark energy} \\
This quadrant is both exotic at the background level and unstable at the perturbative level and the condition $w<-1$ indicates phantom like expansion, while $c_s^2<0$ produces a gradient instability. On sufficiently small scales, perturbations grow approximately as
\begin{equation}
\delta \sim e^{|c_s|kt},
\end{equation}
which signals a rapid breakdown of the linear perturbation description. Note that, if an effective reconstruction inferred $w<-1$ together with $c_s^2<0$, it would not merely indicate phantom evolution. It would also imply that the perturbation sector is unstable within the effective description. Such a result would typically be interpreted either as evidence that the model is pathological, that the inferred regime is only transient, or that the effective fluid description is incomplete and must be replaced by a more fundamental theory in which the apparent instability is resolved.

\item \textbf{$w>-1,\; c_s^2<0$: nonphantom but perturbatively unstable dark energy} \\
This quadrant has a comparatively ordinary background but an unstable perturbation sector. Here since $w>-1$, the background evolution is quintessence like rather than phantom like but $c_s^2<0$ again produces a gradient instability, so perturbations grow rapidly instead of being stabilized by pressure support. Therefore this regime is not generally viable for a sustained dark energy phase, even if its background expansion appears acceptable. Unstable quintessence like behavior can also be seen here, as a model can satisfy nonphantom energy conditions and still fail as a cosmological model because its perturbations are unstable. If observations appeared to favor this region, the conclusion would not simply be that dark energy has unusual clustering behavior. Rather, it would suggest either a transient effective regime, a breakdown of the fluid approximation or the need for additional degrees of freedom that stabilize the perturbations.
\end{itemize}

From an observational and theoretical standpoint, the most natural region for dark energy to occupy is therefore near the boundary
\begin{equation}
w\approx -1,\qquad c_s^2\gtrsim 0
\end{equation}
This includes the cosmological constant limit, nearby quintessence like models, clustering dark energy models with low sound speed and phantom like models that remain perturbatively stable. Current observations strongly favor the vicinity of $w=-1$, but the perturbative coordinate $c_s^2$ remains far less constrained. This is precisely why the $w-c_s^2$ plane is useful: it separates models that are almost identical at the level of the expansion history but very different at the level of perturbations.

Finally, dark energy trajectories need not remain confined to a single quadrant. A crossing of $w=-1$ corresponds to a transition between quintessence like and phantom like behavior, which usually requires multiple degrees of freedom, noncanonical kinetic structure or modified gravity. A crossing of $c_s^2=0$ corresponds to a change in perturbative stability and must be treated with particular care. Thus the trajectory in the $w-c_s^2$ plane encodes not only the instantaneous state of dark energy, but also the possible dynamical evolution of its underlying theory. In this sense, the quadrant in which the trajectory begins, crosses and terminates carries information about both observational signatures and the possible future fate of the Universe.
\begin{figure*}[t]
    \centering
    \includegraphics[width=0.9\linewidth]{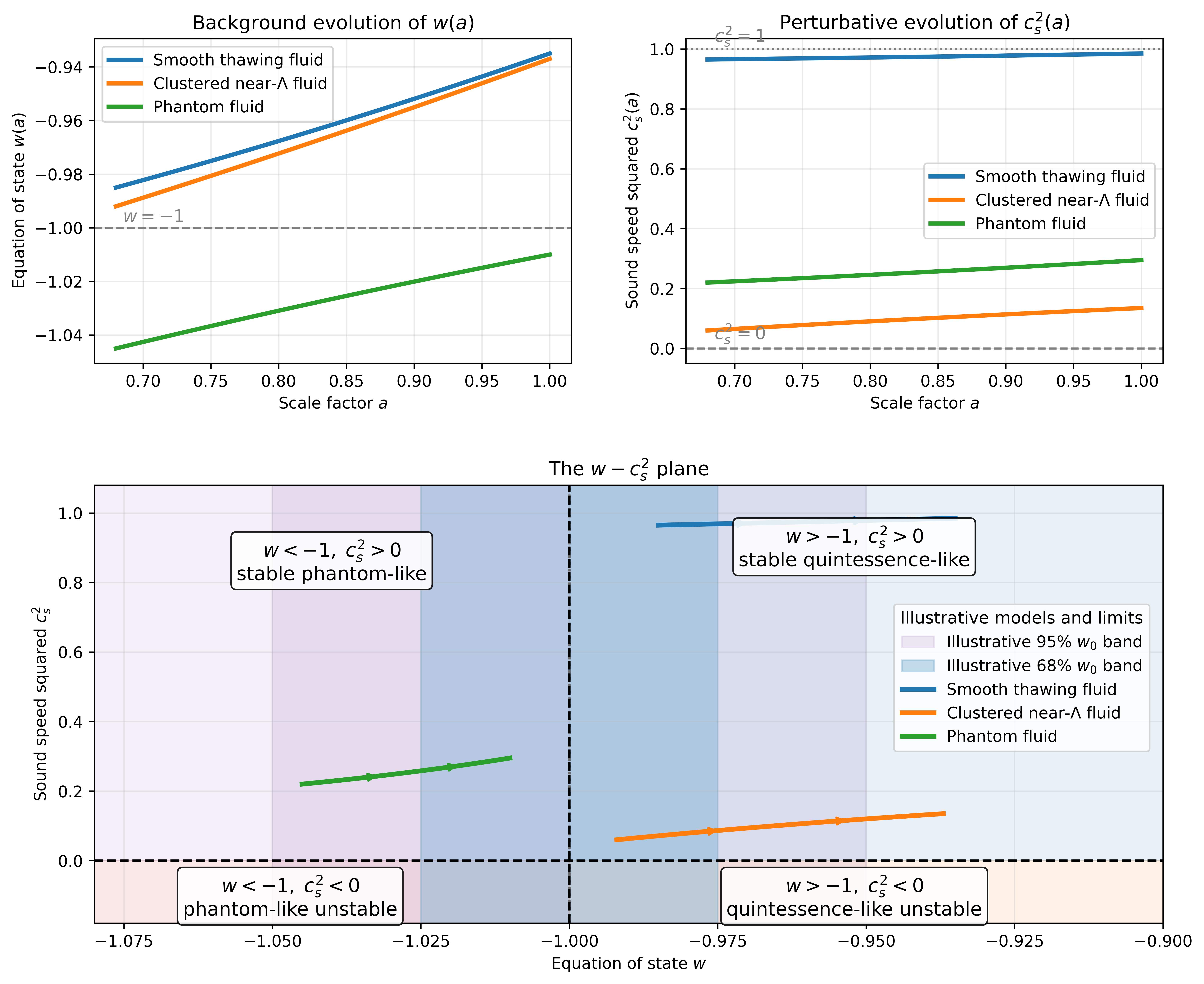}
    \caption{Illustration of the basic $w-c_s^2$ plane, showing the time evolution of $w(a)$ and $c_s^2(a)$ (top panels) and the resulting trajectories in the $w-c_s^2$plane (bottom panel).}
\label{wcs211}
\end{figure*}

We have illustrated the plane in Fig. \ref{wcs211}, where the upper panels show the separate histories of $w(a)$ and $c_s^2(a)$, while the lower panel shows the corresponding trajectories in the $w-c_s^2$ plane. The curves are not generated from a specific scalar field Lagrangian or a particular potential, but are instead phenomenological effective-fluid trajectories chosen to illustrate the classification scheme. In particular, we use a smooth thawing fluid with $w(a)>-1$ and $c_s^2(a)\simeq 1$, a clustered near-$\Lambda$ fluid with $w(a)\simeq -1$ but $c_s^2(a)\ll 1$, and a phantom fluid close to the phantom divide with $w(a)<-1$ and $c_s^2(a)>0$. Schematically, these may be viewed as effective late-time expansions of the form $w(a)=w_0+w_a(1-a)+\cdots$ and $c_s^2(a)=c_{s0}^2+c_{sa}^2(1-a)+\cdots$, rather than as fully specified microscopic models. The grey vertical bands in the lower panel represent an illustrative background constraint centered on $w_0=-1$, with representative $68\%$ and $95\%$ widths, while no comparably tight horizontal band is imposed because current constraints on $c_s^2$ are weak and highly model dependent. Thus the figure is meant to show that present background data already localize the theory space near the phantom divide, whereas the perturbative direction remains much more open. We do not include mock data points here because this figure is intended only to define the geometry and physical quadrants of the plane and mock reconstruction with uncertainties is introduced later when the statistical implementation aspects are discussed.

\subsection{Other Cosmological Relations}
The $w$-$c_s^2$ plane is not merely a taxonomic diagram, because once both quantities are known as functions of redshift, one can construct several cosmologically meaningful relations which are inaccessible using either $w$ or $c_s^2$ alone. The key point here is that $w$ controls the background thermodynamic behavior of dark energy, while $c_s^2$ controls its perturbative response, and only their combination determines whether the dark energy fluid is adiabatic or non-adiabatic, whether it clusters efficiently on a given scale, and how strongly it contributes to the sourcing of gravitational potentials. Here, by “thermodynamic” we mean the macroscopic fluid level behavior of dark energy, which refers to how its pressure, density and energy evolution are related through the equation of state parameter $w$.

Consider first the 
non-adiabaticity of the dark energy sector.
The adiabatic sound speed is
\begin{equation}
c_a^2 \equiv \frac{\dot p_{\rm DE}}{\dot \rho_{\rm DE}},
\end{equation}
where the dot denotes the time derivative.
Using Eq. (\ref{wdef}) we have
\begin{equation}
\dot p_{\rm DE}=\dot w\,\rho_{\rm DE}+w\dot \rho_{\rm DE},
\end{equation}
and combining this with 
the continuity equation
\begin{equation}
\dot\rho_{\rm DE}=-3H(1+w)\rho_{\rm DE},
\end{equation}
we obtain
\begin{equation}
c_a^2
= w - \frac{\dot w}{3H(1+w)}
= w - \frac{a\,w'(a)}{3[1+w(a)]},
\end{equation}
where the prime denotes the derivative with respect to the scale factor $a$.
This quantity is fixed entirely by the background evolution and the physical rest-frame sound speed $c_s^2$; however, $c_s^2$ need not equal $c_a^2$, as is well known. The difference between $c_s^2$ and $c_a^2$ directly measures the intrinsic entropy content of the fluid, and the non-adiabatic pressure perturbation is
\begin{equation}
\delta p_{\rm non-ad}
= \delta p - c_a^2 \delta \rho
= (c_s^2-c_a^2)\,\delta\rho_{\rm rf},
\end{equation}
where $\delta\rho_{\rm rf}$ is the density perturbation in the dark energy rest frame. Equivalently, one may define the intrinsic entropy perturbation through
\begin{equation}
w\,\Gamma_{\rm DE}=(c_s^2-c_a^2)\,\delta_{\rm rf}.
\end{equation}
Thus, once a point or trajectory in the $w$-$c_s^2$ plane is specified, one can immediately determine whether dark energy behaves as a barotropic fluid, for which $c_s^2=c_a^2$, or whether it carries genuine non-adiabatic degrees of freedom, for which $c_s^2\neq c_a^2$. In this sense, the mismatch between the horizontal and vertical coordinates of the diagram has direct thermodynamic significance and one can understand the non-adiabaticity of dark energy as it evolves.

A second quantity of interest is the clustering threshold scale of dark energy. In order to define this scale in a simple analytic way, we work in the Newtonian gauge and consider scalar perturbations around an FLRW background. We assume that the dark energy sector can be treated as an effective fluid with rest frame sound speed $c_s^2$, negligible anisotropic stress and no direct non-gravitational coupling to matter. The estimate below should be concretely understood as a quasistatic, subhorizon approximation, which is valid when $k/a \gg H$ so that time derivatives of the metric potentials and gauge dependent corrections are subleading compared with the pressure gradient and gravitational terms. In this limit, one should understand that the density contrast may equivalently be interpreted as the comoving density perturbation, since gauge differences are suppressed by powers of $aH/k$. With these assumptions, the dark energy perturbation equation can be written as
\begin{equation} \label{gpe}
\ddot\delta_{\rm DE}+(1-3w)H\dot\delta_{\rm DE}
+\left(\frac{c_s^2k^2}{a^2}-4\pi G\rho_{\rm DE}(1+w)\right)\delta_{\rm DE}
\simeq 0,
\end{equation}
where the first term in parentheses represents the pressure gradient contribution and the second term represents the effective gravitational clustering contribution of the dark energy fluid. This equation is not meant to be the full gauge invariant perturbation equation for dark energy on all scales. Rather, it is the Newtonian subhorizon limit which captures the competition between pressure support and gravitational collapse. A more complete treatment would include the velocity perturbation, metric source terms, entropy perturbations and the precise gauge invariant density variable, but these refinements do not change our parametric estimate of the Jeans scale. The transition between pressure-supported behavior and gravitational clustering occurs when the pressure gradient term becomes comparable to the gravitational driving term. Thus, for a stable nonphantom fluid with $1+w>0$ we can define the dark energy Jeans scale by
\begin{equation} \label{jeans1}
\frac{c_s^2k_{J,{\rm DE}}^2}{a^2} \sim 4\pi G\rho_{\rm DE}(1+w),
\end{equation}
which gives
\begin{equation}
k_{J,{\rm DE}}^2(a) \approx \frac{4\pi G a^2 \rho_{\rm DE}(a)(1+w)}{c_s^2}
= \frac{3}{2}\frac{a^2H^2(a)\Omega_{\rm DE}(a)(1+w)}{c_s^2}.
\end{equation}
This expression should be interpreted as an order of magnitude clustering threshold rather than as an exact eigenvalue of the full perturbation system. In particular, if $w<-1$, the factor $1+w$ changes sign and the notion of a standard attractive Jeans scale requires additional care, since phantom-like effective fluids generally require a more complete microphysical realization. In such cases, the above expression is best viewed as identifying the magnitude and parametric dependence of the clustering scale, while the stability and sign of the perturbative response must be analyzed within the underlying theory. The important point is that neither $w$ nor $c_s^2$ alone is sufficient to determine the clustering scale, so the factor $1+w$ controls how strongly the dark energy density perturbation gravitates, while $c_s^2$ controls how efficiently pressure gradients resist clustering. It is therefore natural to define the dimensionless clustering efficiency
\begin{equation}
\mathcal{C}(k,a)\equiv \frac{k_{J,{\rm DE}}^2(a)}{k^2} = \frac{3}{2}\frac{a^2H^2(a)\Omega_{\rm DE}(a)(1+w)}{c_s^2k^2}.
\end{equation}
For $\mathcal{C}\gg 1$, the scale $k$ lies above the dark energy Jeans length and dark energy perturbations are able to cluster efficiently. For $\mathcal{C}\ll 1$, pressure support suppresses the perturbation and the dark energy component remains effectively smooth on that scale. The $w$-$c_s^2$ diagram hence determines not only the instantaneous thermodynamic character of dark energy, but also the range of wavelengths over which it can contribute to cosmic structure.

A third quantity of interest is the effective gravitational response induced by dark energy clustering. In the same Newtonian gauge, subhorizon and quasistatic limit and assuming negligible anisotropic stress so that the two scalar potentials coincide, the Newtonian potential satisfies the Poisson equation
\begin{equation}
k^2\Phi = -4\pi G a^2\left(\rho_m\delta_m+\rho_{\rm DE}\delta_{\rm DE}\right).
\end{equation}
Again $\delta_{\rm DE}$ should be understood as the appropriate rest frame comoving dark energy density perturbation in the limit where gauge corrections are negligible. To make contact with the $w$-$c_s^2$ plane, it is useful to introduce the phenomenological interpolation
\begin{equation}
\delta_{\rm DE}(k,a)\simeq (1+w)\,\frac{1}{1+\left(k/k_{J,{\rm DE}}\right)^2}\,\delta_m(k,a).
\end{equation}
This expression is not intended as an exact solution of the full perturbation equations. It is a compact fitting form which captures the two relevant limits and on scales much larger than the dark energy Jeans scale, $k\ll k_{J,{\rm DE}}$, pressure support is inefficient and the dark energy perturbation can follow the matter perturbation with an amplitude suppressed by $1+w$. On scales much smaller than the Jeans scale, $k\gg k_{J,{\rm DE}}$, the pressure gradient term dominates and dark energy clustering is suppressed as expected. Substituting this interpolating expression into the Poisson equation gives
\begin{equation}
k^2\Phi = -4\pi G a^2 \rho_m \delta_m
\left[ 1+\frac{\rho_{\rm DE}}{\rho_m}(1+w)\frac{1}{1+\left(k/k_{J,{\rm DE}}\right)^2}.
\right]
\end{equation}
This motivates the effective potential response factor
\begin{equation}
\mu_{\rm DE}(k,a)
\equiv
1+\frac{\Omega_{\rm DE}(a)}{\Omega_m(a)}(1+w)
\frac{1}{1+\left(k/k_{J,{\rm DE}}\right)^2},
\end{equation}
which measures the enhancement of the Newtonian potential relative to the case in which dark energy is perfectly smooth. Its dependence on $w$ and $c_s^2$ is straightforward: the factor of $1+w$ determines whether the dark energy fluid carries a clustering degree of freedom at all, while $c_s^2$ determines the scale below which pressure support suppresses that clustering. This indicates that within the quasistatic subhorizon regime, a location or trajectory in the $w$-$c_s^2$ plane maps directly onto a prediction for the scale-dependent contribution of dark energy to the late-time gravitational field. 
 
These three quantities, namely the intrinsic entropy perturbation, the dark energy Jeans scale, and the effective potential response factor, demonstrate that the $w$-$c_s^2$ plane contains much more information than a mere classification of models. Once both coordinates are known, the diagram can be used to reconstruct the thermodynamic character of the fluid, the scale at which it begins to cluster and the extent to which it modifies the sourcing of cosmological potentials. So the plane provides not only a taxonomy of dark energy models but also a compressed representation of their perturbative cosmological consequences.

Note also that this plane naturally provides visualizations and insights into various parameterizations of $w(z)$ by elevating them from purely background descriptions to trajectories in a combined background-perturbation phase space. In standard analyses, parameterizations such as the Chevallier-Polarski-Linder (CPL) are interpreted only through their impact on the expansion history via $H(z)$, but in the present framework any such parametrization maps to a curve
\begin{equation}
z \;\longmapsto\; \bigl(w(z),\,c_s^2(z)\bigr),
\end{equation}
which encodes not only how dark energy evolves in time but also how its perturbative properties co-evolve. This immediately resolves an important degeneracy, which is that different models that share identical $w(z)$ but differ in $c_s^2(z)$ will trace distinct trajectories in this plane, and hence can be discriminated observationally once perturbative information becomes available. Thus, parametrizations of $w(z)$ acquire a richer interpretation when embedded in the $w$-$c_s^2$ plane, as they no longer define a single curve in background space but a family of physically distinct trajectories depending on the associated sound speed behavior. \\

\section{Microphysical Flow Parameter and Stringent DE Tests}
\subsection{Defining the Parameter}
It is not only the location of a given dark energy model in the $w-c_s^2$ plane that is of interest, but also the manner in which the model evolves through this parameter space.  Any dark energy model is characterized by a trajectory
\begin{equation}
a \;\mapsto\; \big(w(a),\, c_s^2(a)\big),
\end{equation}
where $w(a)$ governs the background evolution and $c_s^2(a)$ determines the perturbative response. While the point $(w,c_s^2)$ already encodes valuable information, the local geometry of the trajectory provides further insight into the underlying microphysics. In particular, the slope of the trajectory
\begin{equation}
\mathcal{F}(a) \equiv \frac{d c_s^2}{dw} = \frac{d c_s^2/da}{dw/da} = \frac{\dot{c}_s^2}{\dot{w}},
\end{equation}
defines a microphysical flow parameter which quantifies how the perturbative sector evolves relative to the background evolution. This quantity captures information that is not contained in either $w$ or $c_s^2$ individually and in particular encodes how the structure of the underlying Lagrangian evolves dynamically.

To make this connection explicit, consider a general scalar field realization of dark energy with action
\begin{equation}
S = \int d^4x \sqrt{-g}\, P(\phi, X), \qquad X = -\frac{1}{2} \partial_\mu \phi \partial^\mu \phi,
\end{equation}
for which the energy density and pressure are
\begin{equation}
\rho = 2X P_{,X} - P, \qquad p = P,
\end{equation}
and hence the equation of state parameter is
\begin{equation}
w = \frac{P}{2X P_{,X} - P}.
\end{equation}
The effective sound speed governing perturbations in the rest frame is given by
\begin{equation}
c_s^2 = \frac{P_{,X}}{P_{,X} + 2X P_{,XX}}.
\end{equation}
It is immediately clear that $w$ depends on the ratio of the pressure to the energy density, while $c_s^2$ depends sensitively on the second derivative $P_{,XX}$, which encodes the nonlinear kinetic structure of the theory.  Even though $\mathcal{F}$ encodes how the kinetic structure of the Lagrangian evolves, the determination of $w$, $c_s^2$, and $F$ does not uniquely reconstruct $P(\phi,X)$ but instead heavily restricts the space of allowed theories.

To understand the evolution of these quantities, we differentiate them with respect to time. For the equation of state parameter we find
\begin{equation}
\dot{w} = \frac{\dot{P}(2X P_{,X} - P) - P \frac{d}{dt}(2X P_{,X} - P)}{(2X P_{,X} - P)^2}.
\end{equation}
Expanding the derivatives,
\begin{equation}
\dot{P} = P_{,\phi}\dot{\phi} + P_{,X}\dot{X},
\end{equation}
and
\begin{equation}
\frac{d}{dt}(2X P_{,X}) = 2\dot{X} P_{,X} + 2X (P_{,XX}\dot{X} + P_{,X\phi}\dot{\phi}),
\end{equation}
so that $\dot{w}$ depends on $P_{,\phi}$, $P_{,X}$, $P_{,XX}$ and their evolution through $\dot{\phi}$ and $\dot{X}$. Similarly, for the sound speed
\begin{multline}
\dot{c}_s^2 = \frac{1}{(P_{,X} + 2X P_{,XX})^2} (P_{,XX}\dot{X} + P_{,X\phi}\dot{\phi})(P_{,X} + 2X P_{,XX}) \\ - P_{,X}\left(P_{,XX}\dot{X} + 2X P_{,XXX}\dot{X} + 2X P_{,XX\phi}\dot{\phi}\right) ,
\end{multline}
which explicitly involves higher derivatives such as $P_{,XXX}$. Taking the ratio of these expressions, one obtains
\begin{equation} \label{timede}
\mathcal{F} = \frac{\dot{c}_s^2}{\dot{w}},
\end{equation}
which shows that $\mathcal{F}$ depends on the relative evolution of $P_{,X}$, $P_{,XX}$, and higher derivatives of the Lagrangian. Thus, while $w$ probes the overall energy balance and $c_s^2$ probes the instantaneous kinetic structure, $\mathcal{F}$ probes how this kinetic structure evolves relative to the background dynamics.

The sign of $\mathcal{F}$ carries immediate physical significance, but its interpretation must be tied to the direction in which the background equation of state evolves. A positive value of $\mathcal{F}$ implies that $c_s^2$ and $w$ vary in the same direction, whereas a negative value implies that they vary in opposite directions. Thus, if the background evolves toward more negative values of $w$, as in a freezing evolution toward $w=-1$, then $\mathcal{F}>0$ corresponds to a decreasing sound speed and hence to a reduction in pressure support. In this case the dark energy sector tends to cluster more as it approaches the cosmological constant boundary. On the other hand, for the same freezing evolution, $\mathcal{F}<0$ implies that $c_s^2$ increases as $w$ decreases so that pressure support becomes more efficient and perturbations are increasingly suppressed. For thawing evolution, the interpretation is reversed as in thawing quintessence, the field is initially Hubble damped near $w\simeq -1$ and then evolves away from the cosmological constant limit, so that $w$ typically increases with time. In such a case, $\mathcal{F}<0$ implies that $c_s^2$ decreases as the field thaws, thus signaling reduced pressure support and a possible enhancement of dark energy clustering. On the other hand, $\mathcal{F}>0$ implies that $c_s^2$ increases as $w$ increases, corresponding to a more efficiently pressure supported and smoother perturbative sector. The special case $\mathcal{F}=0$ corresponds to models in which the sound speed is constant along the background trajectory, with canonical quintessence providing the simplest example, since $c_s^2=1$ identically, irrespective of whether the background evolution is thawing or freezing.

The slope $\mathcal{F}$ is only the first level of geometric information encoded in the trajectory and one can go further and consider the curvature
\begin{equation}
\mathcal{K} \equiv \frac{d^2 c_s^2}{dw^2},
\end{equation}
which measures how the slope itself evolves. In terms of time derivatives,
\begin{equation}
\mathcal{K} = \frac{d}{dw}\left(\frac{\dot{c}_s^2}{\dot{w}}\right) = \frac{1}{\dot{w}} \frac{d}{dt}\left(\frac{\dot{c}_s^2}{\dot{w}}\right).
\end{equation}
This quantity depends on second derivatives of $w$ and $c_s^2$, and hence on even higher derivatives of the Lagrangian such as $P_{,XXX}$ and mixed derivatives. Thus, while the point $(w,c_s^2)$ probes the zeroth-order properties of the theory the slope $\mathcal{F}$ probes first-order evolution of the microphysics and the curvature $\mathcal{K}$ probes second-order evolution.

This naturally leads to a hierarchy of observational consistency tests. A viable model must first reproduce the observed values of $(w,c_s^2)$ at a given epoch. It must then reproduce the observed value of $\mathcal{F}$ and ultimately the observed curvature $\mathcal{K}$ if data allow. Each successive condition imposes stronger additional constraints on the allowed functional form of $P(\phi,X)$ and its derivatives and in particular, many models that can reproduce the same $w(a)$ can be distinguished once the evolution of $c_s^2$ is taken into account and even further distinguished once the relation between their evolutions is constrained.

To illustrate this explicitly, consider two models that produce nearly identical background evolution.  This first is canonical quintessence with
\begin{equation}
P_1(\phi,X) = X - V(\phi).
\end{equation}
For canonical quintessence, the sound speed is
\begin{equation}
c_s^2 = 1,
\end{equation}
identically and hence
\begin{equation}
\mathcal{F} = 0, \qquad \mathcal{K} = 0,
\end{equation}
independently of the form of $V(\phi)$. The second model is a a
scalar field with a noncanonical kinetic term,
\begin{equation} \label{kesse}
P_2(\phi,X) = X - V(\phi) + \alpha X^2,
\end{equation}
for which
\begin{equation}
P_{,X} = 1 + 2\alpha X, \qquad P_{,XX} = 2\alpha
\end{equation}
and therefore
\begin{equation}
c_s^2 = \frac{1 + 2\alpha X}{1 + 6\alpha X}.
\end{equation}
In this model, $c_s^2$ evolves with $X$, leading to a time dependent $c_s^2$ even if the background evolution $w(a)$ is tuned to match that of the canonical model. Differentiating gives
\begin{equation}
\dot{c}_s^2 = \frac{(2\alpha \dot{X})(1+6\alpha X) - (1+2\alpha X)(6\alpha \dot{X})}{(1+6\alpha X)^2}
= -\frac{4\alpha \dot{X}}{(1+6\alpha X)^2},
\end{equation}
so that in general,
\begin{equation}
\mathcal{F} = \frac{\dot{c}_s^2}{\dot{w}} \neq 0.
\end{equation}
Even though the two models can be made to agree at the level of $w(a)$, they produce different trajectories in the $w-c_s^2$ plane and hence different values of $\mathcal{F}$ and $\mathcal{K}$. An observational determination of a nonzero $\mathcal{F}$ would therefore immediately rule out the canonical model.

More generally, each class of dark energy models occupies a distinct region not only in $(w,c_s^2)$ but also in the space of allowed trajectories and their derivatives. The requirement that a model reproduce the observed point, slope and curvature imposes increasingly stringent constraints on the derivatives of the Lagrangian, effectively mapping observational data onto constraints on $P_{,X}$, $P_{,XX}$ and higher order terms. Models that require pathological behavior, such as $c_s^2<0$ or rapidly varying higher derivatives to reproduce the observed flow can be excluded on theoretical grounds. 
\begin{figure*}[t]
    \centering
    \includegraphics[width=0.9\linewidth]{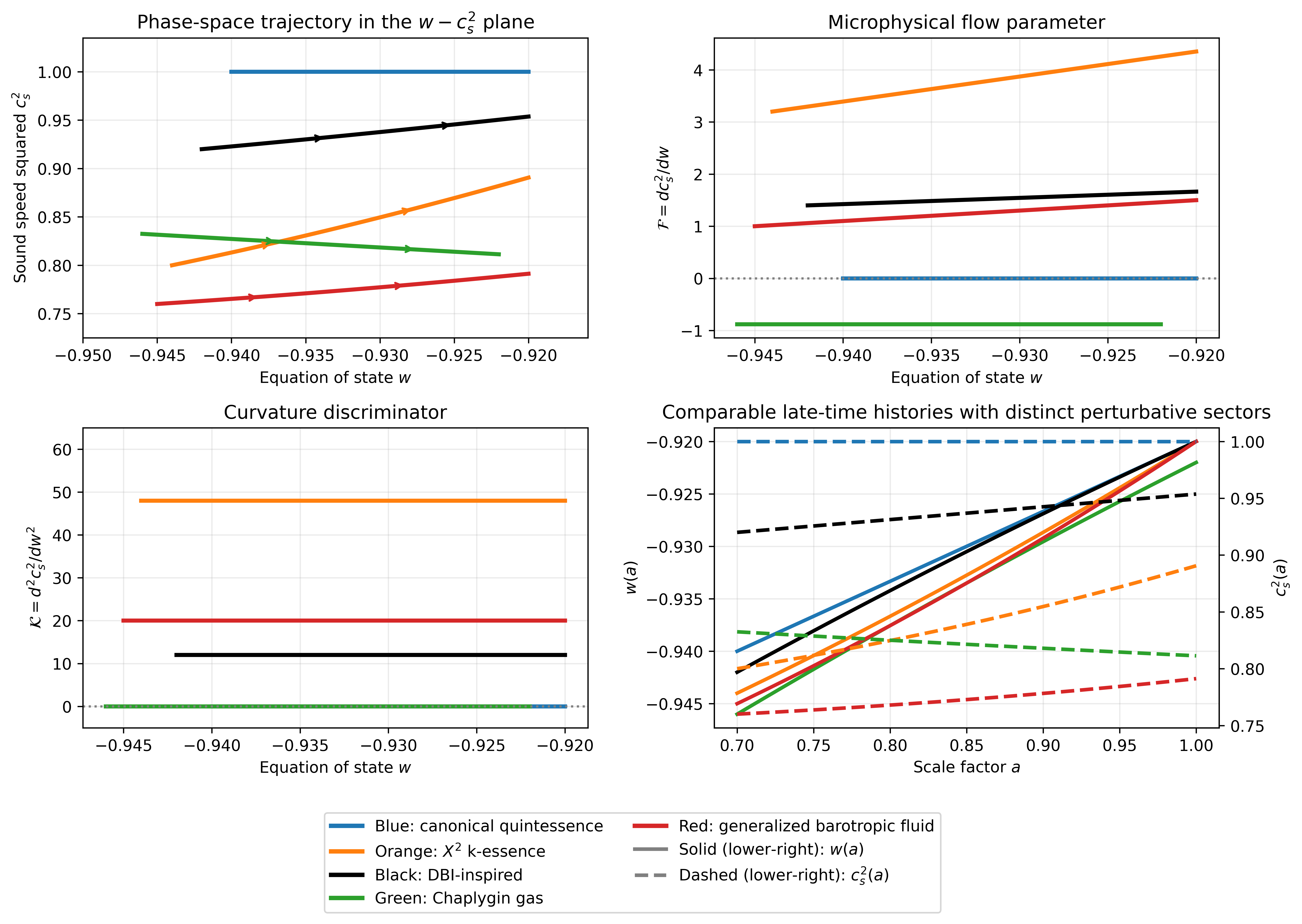}
    \caption{Illustrative navigation of the dark energy model zoo using the $w-c_s^2$ plane,
    $\mathcal{F}=d c_s^2/dw$ and $\mathcal{K}=d^2 c_s^2/dw^2$}.
    \label{micro}
\end{figure*}
We have illustrated the utility of the microphysical flow parameter, the derivative hierarchy, and the $w-c_s^2$ plane in Fig.~\ref{micro}. The models shown in the figure are chosen so that their late time EOS histories $w(a)$ are deliberately close to one another, which would show them mimicking the common situation in which different dark energy scenarios are nearly degenerate at the level of background expansion. The first reference case is canonical quintessence, described by a Lagrangian of the form $P(\phi,X)=X-V(\phi)$. Beyond this, we can also explore barotropic models where the conventional example is the Chaplygin gas \cite{cp1Kamenshchik:2001cp,cp2Bento:2002ps,cp3Bento:2004uh,cp4Sandvik:2002jz}, which provides a useful contrasting example because its pressure is specified directly as a function of the density. In the generalized Chaplygin gas case one usually writes
\begin{equation}
p=-\frac{A}{\rho^\alpha}
\end{equation}
so that
\begin{equation}
w=\frac{p}{\rho}=-\frac{A}{\rho^{1+\alpha}}
\end{equation}
and
\begin{equation}
c_s^2=\frac{dp}{d\rho}=-\alpha w
\end{equation}
This produces a simple, approximately linear relation between $c_s^2$ and $w$ in the phase space so that unlike canonical quintessence, the Chaplygin gas does not sit at fixed sound speed. Its perturbative response is still tightly tied to its background equation of state. We also consider another barotropic fluid example in which the pressure is specified as a more general function of the density, $p=p(\rho)$. In such cases,
\begin{equation}
w=\frac{p(\rho)}{\rho}
\end{equation}
while
\begin{equation}
c_s^2=\frac{dp}{d\rho}
\end{equation}
so the relation between $w$ and $c_s^2$ depends on the functional form of $p(\rho)$. For the illustrative trajectory shown in the figure, we use a local nonlinear barotropic relation between the sound speed and the equation of state, of the schematic form $c_s^2(w)=c_*^2+\alpha(w-w_*)+\beta(w-w_*)^2$, with parameters chosen so that the model has a late-time $w(a)$ history close to the other examples but a distinct perturbative response. This should be understood as a phenomenological barotropic expansion around the observed late-time regime, rather than as the Chaplygin relation $c_s^2=-\alpha w$. A nonlinear barotropic relation hence produces a trajectory that can differ from both the Chaplygin gas and scalar field examples, even if its late-time $w(a)$ is arranged to be very similar. Another interesting representative class consists of noncanonical scalar field models, for which we consider an $X^2$ k-essence deformation by a kinetic structure of the form in Eq. (\ref{kesse}). Similarly, a DBI-inspired scalar field has a noncanonical kinetic structure, often represented schematically by
\begin{equation}
P(\phi,X)=-\frac{1}{f(\phi)}\left(\sqrt{1-2f(\phi)X}-1\right)-V(\phi),
\end{equation}
with a sound speed of the form
\begin{equation}
c_s^2=\frac{1}{\gamma^2}=1-2f(\phi)X,
\end{equation}
where $\gamma$ is the usual DBI Lorentz factor. This naturally generates a curved trajectory in the $w-c_s^2$ plane because the perturbative response depends on the kinetic structure and warp factor, not only on the background equation of state. 

The role of Fig.~\ref{micro} is to show how these differences become visible once the model space is viewed through the combined background-perturbation phase space. The bottom right panel demonstrates that the chosen models can have nearly indistinguishable late time histories $w(a)$, which means that expansion data alone would have difficulty separating them. The top left panel then shows that the same models trace distinct paths in the $w-c_s^2$ plane, showing differences in their perturbative sectors. The top right panel further sharpens the difference by plotting the microphysical flow parameter which separates models with fixed sound speed from those in which the kinetic or fluid microphysics evolves along with the background. The bottom left panel then displays the curvature discriminator which isolates genuinely nonlinear trajectories in the phase space. Models that may look similar at the level of $\mathcal{F}$ can therefore still be distinguished once $\mathcal{K}$ is considered and a subtle but important point is that the numerical values of $\mathcal{F}$ and $\mathcal{K}$ can appear large even though $c_s^2$ itself remains bounded between $0$ and $1$. This happens because the derivatives are taken with respect to $w$, whose late-time variation is confined to a narrow interval. Thus, the figure demonstrates that even when dark energy models are highly degenerate at the background level, their trajectories and higher order geometric structure in the $w-c_s^2$ plane provide a robust hierarchical discriminator of the underlying microphysics. 

\subsection{EFT connections}
The microphysical flow parameter admits a natural interpretation within the effective field theory of dark energy \cite{eft1gubitosi2013effective,eft2linder2016effective,eft3bloomfield2013dark,eft4liang2023dark,eft5frusciante2014effective}, where both the background evolution and perturbative dynamics are governed by a set of time dependent operator coefficients. In this framework, we see that the scalar sector is typically characterized by functions such as $\alpha_K$, $\alpha_B$, $\alpha_M$ and $\alpha_T$, together with the background expansion rate $H(a)$ and its derivatives. The dark energy equation of state $w$ and the sound speed $c_s^2$ are not fundamental parameters, but derived quantities constructed from these EFT functions. At the background level, the equation of state is determined by the effective energy density and pressure of the dark sector written in the usual way:
\begin{equation*}
w = \frac{p_{\rm DE}}{\rho_{\rm DE}},
\end{equation*}
where $\rho_{\rm DE}$ and $p_{\rm DE}$ depend on the background EFT functions and the Hubble rate. At the perturbative level, the scalar sound speed can be written schematically as
\begin{equation}
c_s^2 = c_s^2\big(\alpha_K,\alpha_B,\alpha_M,\alpha_T,H,\dot{H},\ldots\big),
\end{equation}
with explicit expressions depending on the specific EFT realization. The sound speed here encodes information about the kinetic structure and mixing of scalar degrees of freedom. The flow parameter $\mathcal{F}$ can be expressed as a ratio of time derivatives as in \eqref{timede} and using the chain rule, we can hence write
\begin{equation}
\dot{c}_s^2 = \sum_I \frac{\partial c_s^2}{\partial \alpha_I}\dot{\alpha}_I
+ \frac{\partial c_s^2}{\partial H}\dot{H}
+ \frac{\partial c_s^2}{\partial \dot{H}}\ddot{H} + \cdots
\end{equation}
\begin{equation}
\dot{w} = \sum_I \frac{\partial w}{\partial \alpha_I}\dot{\alpha}_I
+ \frac{\partial w}{\partial H}\dot{H}
+ \frac{\partial w}{\partial \dot{H}}\ddot{H} + \cdots
\end{equation}
where the index $I$ runs over the relevant EFT functions and combining these expressions gives us
\begin{equation}
\mathcal{F} = \frac{ \sum_I (\partial c_s^2/\partial \alpha_I)\dot{\alpha}_I + (\partial c_s^2/\partial H)\dot{H} + \cdots}{
\sum_I (\partial w/\partial \alpha_I)\dot{\alpha}_I + (\partial w/\partial H)\dot{H} + \cdots
}
\end{equation}
This form makes explicit that $\mathcal{F}$ is not associated with any single EFT coefficient, but instead probes the relative evolution of the perturbative and background sectors. In particular, it is sensitive to the time dependence of the EFT functions and therefore has information about the dynamical structure of the underlying dark energy action. Canonical scalar field models, for which $c_s^2 = 1$ identically, trivially would give $\mathcal{F} = 0$ while more general constructions such as k-essence or models with nontrivial braiding can generate nonzero flow through the time evolution of the kinetic operators. This gives $\mathcal{F}$ a distinctive role, which is that once a background theory has been specified, a measurement of $\mathcal{F}$ constrains not only the trajectory $c_s^2(w)$ but also the allowed time evolution of the action level functions $\alpha_I(t)$. In this sense, $\mathcal{F}$ acts as a direct bridge between phenomenological observables and the kinetic, braiding, and higher derivative operators that define the underlying dark energy theory. Note that an interesting possibility, for future cosmological studies, could be to investigate under what conditions does $\mathcal{F}$ become singular. From this perspective, $\mathcal{F}$ can be viewed as a derived observable that compresses the information contained in the EFT functions into a single quantity that directly links background evolution to perturbative dynamics. It thus provides a useful bridge between observational diagnostics and the microphysical structure of the dark sector.

\subsection{Microphysical Flow, $H_0$ and $\sigma_8$}
We begin by deriving a relation between the microphysical flow parameter
\begin{equation}
\mathcal{F}(z) \equiv \frac{d c_s^2}{dw},
\end{equation}
and the observables $H_0$ and $\sigma_8$, with the key point being that $H_0$ is primarily constrained by background expansion observables, while $\sigma_8$ is sensitive to both background evolution and perturbative dynamics. The parameter $\mathcal{F}$ therefore acts as a bridge that determines how changes in the background propagate into the perturbative sector. We begin with the late time Friedmann equation,
\begin{equation}
H^2(z) = H_0^2 \left[ \Omega_m (1+z)^3 + \Omega_{\rm DE}\, e^{3\int_0^z \frac{1+w(z')}{1+z'}dz'} \right].
\label{friedmann_general}
\end{equation}
For clarity, we consider the case of a constant equation of state parameter $w$ for which
\begin{equation}
H^2(z) = H_0^2 \left[ \Omega_m (1+z)^3 + \Omega_{\rm DE} (1+z)^{3(1+w)} \right].
\label{friedmann_constantw}
\end{equation}
Late time cosmological probes such as supernovae and baryon acoustic oscillations constrain the comoving distance,
\begin{equation}
D(z) = \int_0^z \frac{dz'}{H(z')}.
\label{distance_def}
\end{equation}
This can be written schematically as
\begin{equation}
D(z) = \frac{1}{H_0} f(z; w),
\label{distance_scaling}
\end{equation}
where $f(z; w)$ encodes the dependence on the expansion history.  Let us assume that $D(z)$ is fixed by observations, so that any simultaneous change in $H_0$ and $w$ must satisfy
\begin{equation}
\delta D(z) = 0.
\end{equation}
Using Eq. (\ref{distance_scaling}), this would imply to us that
\begin{equation}
0 = \delta\left(\frac{1}{H_0} f(z; w)\right)
= -\frac{f}{H_0^2}\,\delta H_0 + \frac{1}{H_0} \frac{\partial f}.{\partial w}\,\delta w
\end{equation}
Rearranging, we obtain the relation
\begin{equation}
\delta H_0 = \frac{H_0}{f}\frac{\partial f}{\partial w}\,\delta w.
\label{H0_w_relation}
\end{equation}
Equation (\ref{H0_w_relation}) expresses the degeneracy between $H_0$ and $w$ that arises from distance measurements. It shows us that a shift in $w$ must be accompanied by a corresponding shift in $H_0$ in order to preserve the observed distance-redshift relation.

We now consider the impact on structure formation, for which we know that the linear growth of matter perturbations from \eqref{gpe} and \eqref{jeans1} obeys
\begin{equation}
\ddot{\delta}_m + 2H \dot{\delta}_m - 4\pi G_{\rm eff}(a,k)\,\rho_m \delta_m = 0,
\label{growth_eq}
\end{equation}
where $G_{\rm eff}$ encodes the effect of dark energy perturbations. In the presence of clustering dark energy, this can be written schematically as
\begin{equation}
G_{\rm eff}(k,a) \simeq G \left[ 1 + \frac{\Omega_{\rm DE}(a)}{\Omega_m(a)} (1+w)\,\frac{1}{1 + (k/k_J)^2} \right],
\label{geff}
\end{equation}
where the dark energy Jeans scale satisfies
\begin{equation}
k_J^2(a) \sim \frac{(1+w)}{c_s^2}.
\label{jeans_scale}
\end{equation}
The growth amplitude at the present epoch is characterized by $\sigma_8 \propto D(a=1)$, where $D(a)$ is the linear growth factor. To leading order, variations in $\sigma_8$ can be expressed as
\begin{equation}
\delta \ln \sigma_8 \simeq A\,\delta w + B\,\delta c_s^2,
\label{sigma8_variation}
\end{equation}
where the coefficients $A$ and $B$ depend on the background cosmology, the redshift range of the data, the normalization convention for the growth factor, and the scale under consideration. This expression should be understood as a general linear response relation rather than as a model-specific result. This is because around any fiducial dark energy cosmology, the observable $\sigma_8$ responds to small changes in the background expansion history and to small changes in the perturbative properties of the dark energy sector. Since $w$ controls the homogeneous energy density evolution and hence the Hubble friction term in the growth equation, its variation contributes through the first term and since $c_s^2$ controls the ability of dark energy to cluster and modify the gravitational potentials, its variation contributes through the second term. The coefficients $A$ and $B$ therefore tell us the response kernels of the growth observable to these two sectors. Their precise numerical values would depend on the fiducial model and on the dataset being considered, but the structure of Eq. (\ref{sigma8_variation}) is much more general. It simply states tell us that, to first order in departures from a reference cosmology, the change in the late time clustering amplitude can be decomposed into a background contribution and a perturbative contribution. Higher order corrections would involve terms such as $\delta w^2$, ${\delta c_s^4}$ and mixed products, but these are subleading for sufficiently small deviations, so Eq. (\ref{sigma8_variation}) provides a model-independent local expansion of the growth response in the two-dimensional dark energy parameter space spanned by $w$ and $c_s^2$. The microphysical flow parameter relates variations in $c_s^2$ to variations in $w$,
\begin{equation}
\delta c_s^2 \simeq \mathcal{F}\,\delta w.
\label{flow_relation}
\end{equation}
Substituting Eq.(\ref{flow_relation}) into Eq.(\ref{sigma8_variation}), we obtain
\begin{equation}
\delta \ln \sigma_8 \simeq \left(A + B\,\mathcal{F}\right)\delta w.
\label{sigma8_flow}
\end{equation}
Combining Eq.(\ref{sigma8_flow}) with the relation between $\delta H_0$ and $\delta w$ in \eqref{H0_w_relation}), we eliminate $\delta w$ and arrive at
\begin{equation}
\delta \ln \sigma_8 \simeq \frac{A + B\,\mathcal{F}}{C}\,\delta H_0,
\qquad
C \equiv \frac{H_0}{f}\frac{\partial f}{\partial w}.
\label{final_relation}
\end{equation}
Equation (\ref{final_relation}) is the central result here, because it shows that the correlation between $H_0$ and $\sigma_8$ is controlled by the combination $A + B\,\mathcal{F}$. In particular, the microphysical flow parameter $\mathcal{F}$ determines how a change in the background expansion, as encoded in $H_0$, propagates into the growth of structure and this result has an immediate implication. For a given shift in $H_0$, the corresponding shift in $\sigma_8$ is not fixed solely by the background equation of state, but depends on the microphysical flow parameter. In the special case $\mathcal{F}=0$ corresponding to canonical quintessence with $c_s^2=1$, the relation reduces to the standard background-driven correlation. But for $\mathcal{F}\neq 0$, the growth response is modified, allowing for a decoupling between the background and perturbative sectors. 

It follows that $\mathcal{F}_0$, which is the present day value of the microphysical flow parameter, plays a central role in determining whether modifications to the expansion history can be reconciled with observations of large scale structure. In this sense, $\mathcal{F}_0$ is not merely a derived quantity, but a parameter that controls the mapping between background and perturbative cosmology. Any consistent description of the late time Universe must therefore specify not only $w(z)$ and $c_s^2(z)$ individually, but also their relation through the flow parameter.

\section{Measuring $\mathcal{F}_{0}$}
How can we actually determine $\mathcal{F}_0$? What does it mean to determine $\mathcal{F}_0$ from early and late time probes and how can such a quantity be operationally reconstructed from data? The answer is that $\mathcal{F}_0$ is not measured directly as a primary observable, but inferred from a joint reconstruction of the present day values and local redshift dependence of the dark energy equation of state $w(z)$ and the sound speed $c_s^2(z)$. By definition,
\begin{equation}
\mathcal{F}_0 \equiv \left.\frac{dc_s^2}{dw}\right|_{z=0}
= \left.\frac{\dfrac{dc_s^2}{dz}}{\dfrac{dw}{dz}}\right|_{z=0},
\label{F0_def_z}
\end{equation}
provided $dw/dz \neq 0$ in the relevant neighborhood of $z=0$ and equivalently, in terms of the scale factor
\begin{equation}
\mathcal{F}_0
= \left.\frac{\dfrac{dc_s^2}{da}}{\dfrac{dw}{da}}\right|_{a=1}.
\label{F0_def_a}
\end{equation}
This makes clear that the determination of $\mathcal{F}_0$ requires not merely estimates of $w_0$ and $c_{s0}^2$, but a local reconstruction of the trajectory in the $w-c_s^2$ plane near the present epoch.

At the level of data analysis, one may proceed by introducing expansions for both functions where for instance, one may write
\begin{equation}
w(a) = w_0 + w_a(1-a) + \cdots
\label{w_expansion}
\end{equation}
and similarly
\begin{equation}
c_s^2(a) = c_{s0}^2 + c_{sa}^2(1-a) + \cdots,
\label{cs_expansion}
\end{equation}
in which case one immediately finds
\begin{equation}
\mathcal{F}_0 = \frac{c_{sa}^2}{w_a}.
\label{F0_parametric}
\end{equation}
This parametrized form is useful because it shows that $\mathcal{F}_0$ is not an independent background parameter, but rather measures the relative rate at which the perturbative sector changes compared with the background sector near the present epoch. In other words, $w_a$ controls the local departure of the equation of state from its present value, while $c_{sa}^2$ controls the corresponding local evolution of the sound speed. Their ratio therefore determines the local direction of motion in the $w-c_s^2$ plane. More generally, if one reconstructs $w(z)$ and $c_s^2(z)$ in bins or through Gaussian processes, then $\mathcal{F}_0$ is obtained by taking the local derivative ratio at $z=0$ according to Eq.  \eqref{F0_def_z}. In such a reconstruction, the important object is not just the value of $w$ or $c_s^2$ in the lowest redshift bin, but it is instead the local slope of the reconstructed curve connecting nearby points in the phase space. This also means that the extraction of $\mathcal{F}_0$ is more sensitive to the assumed smoothness of the reconstruction than the extraction of $w_0$ or $c_{s0}^2$ individually and the parametric form in Eq. (\ref{F0_parametric}) is hence merely the simplest illustration of a more general point, namely that $\mathcal{F}_0$ is extracted from the local tangent to the reconstructed trajectory in the $w-c_s^2$ plane.

The determination of $w(z)$ and $c_s^2(z)$, however, relies on different classes of observables. The equation of state parameter is principally constrained by background probes and these include type Ia supernova luminosity distances, baryon acoustic oscillation measurements, cosmic chronometers and in combination with a calibration, the local or inverse distance ladder determination of $H_0$. In such analyses, one fits the expansion history
\begin{equation}
H^2(z) = H_0^2 \left[\Omega_m(1+z)^3 + \Omega_{\rm DE}\exp\left(3\int_0^z \frac{1+w(z')}{1+z'}\,dz'\right)\right],
\label{Hz_again}
\end{equation}
to distance and expansion rate data, which produces a posterior distribution for $w(z)$ near $z=0$.

By contrast, $c_s^2(z)$ is not a background quantity and must be inferred from perturbation-sensitive observables. The relevant information enters through the evolution of metric and matter perturbations with Eqs. \eqref{geff} and \eqref{growth_eq} being central to it. Hence the sound speed controls the scale dependence and efficiency of dark energy clustering, and can therefore be constrained through more non-trivial methods hopefully from missions like LILA or through other aspects of large scale structure, CMB lensing, etc. 

Therefore, the phrase ``determining $\mathcal{F}_0$ from late time probes" means reconstructing both $w(z)$ and $c_s^2(z)$ using observables whose dominant sensitivity lies in the low redshift universe, and then evaluating Eqs. (\ref{F0_def_z}) or (\ref{F0_def_a}) at the present epoch. Concretely, one may imagine combining supernovae, BAO, cosmic chronometers, weak lensing, redshift-space distortions and low redshift large scale structure datasets to obtain a late time posterior
\begin{equation}
\mathcal{P}_{\rm late}\big(w_0,w'_0,c_{s0}^2,(c_s^2)'_0,\ldots\big),
\end{equation}
from which one constructs
\begin{equation}
\mathcal{F}_0^{\rm late}
= \frac{(c_s^2)'_0}{w'_0}.
\label{F0_late}
\end{equation}
Here a prime can denote differentiation with respect to either $z$ or $a$, provided the convention is used consistently.

When we talk of early-time probes for determining the microphysical flow parameter, a clarification is necessary since dark energy is subdominant at early times in standard cosmology. What is really meant is not that one literally measures the present day flow in the early Universe, but rather that one infers the value of the late time parameters from early universe data under a given cosmological model. For example, CMB anisotropies, CMB lensing and early time standard ruler information constrain the integrated expansion history and the growth of perturbations across cosmic time. Once a model for $w(z)$ and $c_s^2(z)$ is specified, early time data would induce a posterior on the late time parameters through evolution from recombination to the present and one may therefore define
\begin{equation}
\mathcal{P}_{\rm early}\big(w_0,w'_0,c_{s0}^2,(c_s^2)'_0,\ldots\big),
\end{equation}
and from this obtain
\begin{equation}
\mathcal{F}_0^{\rm early}
= \frac{(c_s^2)'_0}{w'_0}.
\label{F0_early}
\end{equation}
The distinction is therefore conceptual: late time probes reconstruct $\mathcal{F}_0$ more directly from low-redshift expansion and clustering, whereas early time probes reconstruct $\mathcal{F}_0$ indirectly by constraining the present day parameters through a global cosmological fit.

This point may also be formulated in terms of separate likelihoods, wherein if one denotes the late time and early time likelihoods by $\mathcal{L}_{\rm late}$ and $\mathcal{L}_{\rm early}$, then one may write
\begin{equation}
\mathcal{L}_{\rm late} = \mathcal{L}_{\rm late}\big[w(z),c_s^2(z),\Theta\big],
\qquad
\mathcal{L}_{\rm early} = \mathcal{L}_{\rm early}\big[w(z),c_s^2(z),\Theta\big],
\end{equation}
where $\Theta$ denotes the remaining cosmological parameters. Marginalizing these likelihoods over $\Theta$ would give us two posterior distributions for the reconstructed trajectory in the $w-c_s^2$ plane and therefore two posteriors for its present day tangent,
\begin{equation}
P_{\rm late}(\mathcal{F}_0), \qquad P_{\rm early}(\mathcal{F}_0).
\label{F0_posteriors}
\end{equation}
A comparison of these two distributions then provides a direct test of whether the inferred microphysical flow is consistent across epochs and across classes of observables.

There are also important technical subtleties in the determination of $\mathcal{F}_0$ which should also be noted here. First, since $\mathcal{F}_0$ is a derivative ratio it is more sensitive to noise than either $w_0$ or $c_{s0}^2$ individually and so it is therefore perhaps advantageous to determine it from a smooth reconstruction rather than from raw finite differencing of binned data. Second, if $w'_0$ is very small the ratio in Eq. (\ref{F0_def_z}) can become numerically unstable, in which case one may instead define an averaged flow parameter over a narrow redshift interval
\begin{equation}
\bar{\mathcal{F}}_{[0,z_*]}
\equiv
\frac{\Delta c_s^2}{\Delta w}
=
\frac{c_s^2(z_*)-c_{s0}^2}{w(z_*)-w_0},
\label{Fbar}
\end{equation}
which reduces to $\mathcal{F}_0$ in the limit of sufficiently small $z_*$. Third, because $c_s^2$ may vary over orders of magnitude one may also consider a logarithmic flow estimator,
\begin{equation}
\mathcal{F}_{0,\ln}
\equiv
\left.\frac{d\ln c_s^2}{dw}\right|_{z=0},
\label{Flog}
\end{equation}
though the basic logic remains unchanged.

An important practical point is that the reconstruction described above should not be interpreted as a direct algebraic extraction of $\mathcal{F}_0$ from data. In an actual observational analysis, $\mathcal{F}_0$ would have to be treated as a parameter of the cosmological model and propagated through the full Einstein-Boltzmann system. A convenient way to do this would be to choose a local parametrization of the trajectory in the $w-c_s^2$ plane, for example by writing the sound speed near the present epoch as a function of $w$ rather than as an independent function of redshift,
\begin{equation}
c_s^2(a)=c_{s0}^2+\mathcal{F}_0\left[w(a)-w_0\right]+\cdots
\end{equation}
or equivalently, in the linear expansion discussed above, by imposing $c_{sa}^2=\mathcal{F}_0 w_a$. The parameter set sampled in a likelihood analysis would then be enlarged to
\begin{equation}
\Theta_{\rm ext}=\{\Omega_bh^2,\Omega_ch^2,H_0,A_s,n_s,\tau,\ldots,w_0,w_a,c_{s0}^2,\mathcal{F}_0,\ldots\},
\end{equation}
with the dark energy perturbations evolved self consistently for each point in this parameter space. The role of the Boltzmann calculation is crucial here because $\mathcal{F}_0$ does not affect observables directly, but only through its effect on the time and scale dependence of perturbations. For each sampled value of $\Theta_{\rm ext}$, one must solve the coupled background and perturbation equations, obtain the relevant transfer functions and then compute observables such as CMB anisotropies, CMB lensing, matter power spectra, weak lensing spectra, redshift space distortion observables or any future direct probe of large scale gravitational potentials. Schematically, the theoretical prediction entering the likelihood has the form
\begin{equation}
\mathbf{d}_{\rm th}=\mathbf{d}_{\rm th}\left[\Theta_{\rm ext}\right],
\end{equation}
where the dependence on $\mathcal{F}_0$ is encoded through the evolution of $c_s^2(a)$ and so through the dark energy perturbation sector. Thus, the statistically meaningful constraint on $\mathcal{F}_0$ is obtained only after marginalizing over the remaining cosmological and nuisance parameters,
\begin{equation}
P(\mathcal{F}_0|\mathbf{d})
\propto
\int d\Theta_{\rm ext}^{\prime}\,
\mathcal{L}\left(\mathbf{d}\,|\,\Theta_{\rm ext}^{\prime},\mathcal{F}_0\right)
\pi\left(\Theta_{\rm ext}^{\prime},\mathcal{F}_0\right),
\end{equation}
where $\Theta_{\rm ext}^{\prime}$ denotes all parameters except $\mathcal{F}_0$, $\mathcal{L}$ is the full data likelihood and $\pi$ denotes the prior. The derivative ratio expressions above should therefore be understood as defining the geometrical meaning of $\mathcal{F}_0$, while a real measurement of it requires a full Boltzmann plus likelihood analysis in which $\mathcal{F}_0$ is varied on the same footing as the standard cosmological and dark energy parameters.

To illustrate how $\mathcal{F}_0$ can play an important role as a fundamental parameter in the age of cosmic tensions, given that we have proper non-circular measurements of the sound speed in the future, consider the following scenario. Suppose that two dark energy models are found to be indistinguishable at the level of the usual present-day cosmological quantities, such as $H_0$, $w_0$, $c_{s0}^2$ and $\sigma_{8,0}$. In a conventional analysis, this would make the two models appear nearly degenerate, because they would share the same present day expansion rate, the same background equation of state, the same instantaneous sound speed and the same amplitude of matter clustering. However, these quantities only specify the position of the model at the present epoch and they do not specify the local direction in which the model is moving in the $w-c_s^2$ plane. The role of $\mathcal{F}_0$ is precisely to provide this missing information. For concreteness, we consider two simple representative models, wherein model A is taken to have an approximately constant sound speed near the present epoch, so that $c_s^2(z)=c_{s0}^2$ and hence $\mathcal{F}(z)=0$. Model B is chosen to have the same present day values of $H_0$, $w_0$, $c_{s0}^2$ and $\sigma_{8,0}$, but a nontrivial redshift evolution of the sound speed, giving a nonzero present day flow $\mathcal{F}_0$. Both models are assigned the CPL background equation of state evolution, which is defined by
\begin{equation}
w(a)=w_0+w_a(1-a),
\end{equation}
and this is what gives us the same expansion history $H(z)$. The only difference between the models is the local evolution of the perturbative sector encoded in $c_s^2(z)$ and more sharply, in $\mathcal{F}(z)$.
\begin{figure*}[t]
    \centering
    \includegraphics[width=0.9\linewidth]{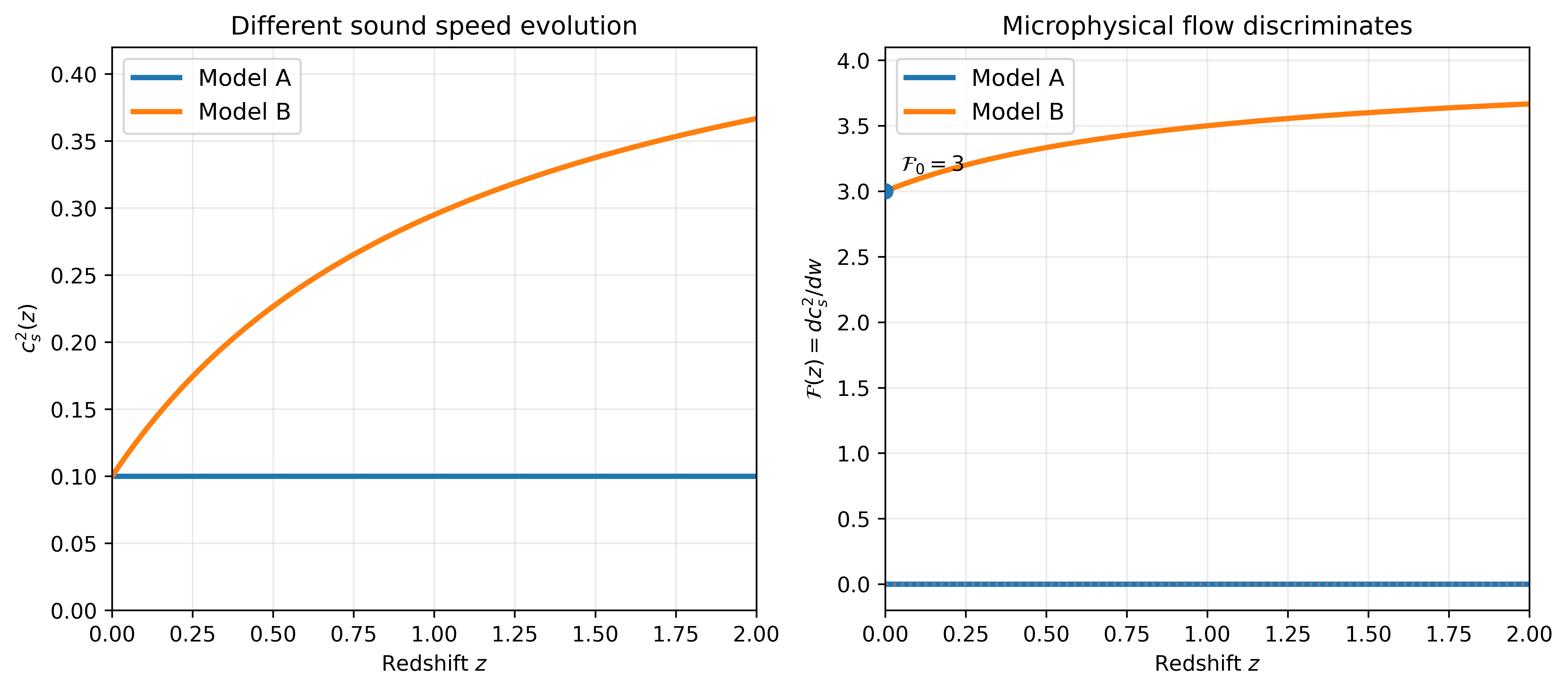}
    \caption{Illustration of two dark energy models with identical expansion history and present day $c_{s0}^2$, but distinct sound speed evolution and therefore distinct microphysical flow $\mathcal{F}(z)$, making $\mathcal{F}_0$ a present epoch discriminator.}
    \label{refplot1}
\end{figure*}
This is shown explicitly in Fig. \ref{refplot1}, wherein it is shown that even though both models have the same value of $c_s^2$ at $z=0$, their sound speed histories differ away from the present epoch, as shown in the left panel. The right panel then displays the key diagnostic quantity, $\mathcal{F}(z)$, and we see that for model A, $\mathcal{F}(z)=0$ while Model B has a nonzero flow, with a finite value $\mathcal{F}_0$ at the present epoch. So, even though the two models agree in their instantaneous present day values, they are separated by the local slope of their trajectory in the $w-c_s^2$ plane. 

The important lesson from Fig. \ref{refplot1} is that $\mathcal{F}_0$ is not merely another way of parametrizing $w_0$ or $c_{s0}^2$. Instead, it is something that captures information about the local motion of the dark energy model in the combined background perturbation phase space. In the context of future cosmological tensions, this is especially relevant because if two cosmological descriptions agree on the usual scalar quantities but differ in the relation between background evolution and perturbative response, then the tension is not visible in $H_0$, $w_0$, $c_{s0}^2$ or $\sigma_{8,0}$ separately. It appears instead in the flow structure connecting these sectors and so, provided future observations can deliver independent and non-circular constraints on $c_s^2(z)$, the parameter $\mathcal{F}_0$ can become a powerful diagnostic of whether apparently degenerate cosmological models are truly microphysically equivalent. \\

\section{Conclusions}
In this work, we have developed a unified framework for probing the nature of dark energy by extending traditional background diagnostics into a combined background-perturbation phase space. We have introduced the $w$-$c_s^2$ plane as a minimal yet physically complete representation of dark energy at linear order, where the equation of state $w$ captures the expansion history and the sound speed $c_s^2$ governs the perturbative response. Within this framework, dark energy models are naturally described as trajectories rather than isolated points, thus allowing for a transparent classification of their dynamical and clustering behavior. We have shown that this plane encodes not only a taxonomy of models, but also direct access to thermodynamic properties, clustering scales and gravitational response, thereby providing a compact yet information rich description of dark energy phenomenology.

Building on this phase-space picture, we introduced the microphysical flow parameter $\mathcal{F} = dc_s^2/dw$, which quantifies how the perturbative sector evolves relative to the background dynamics. We demonstrated that $\mathcal{F}$ is sensitive to the kinetic structure of the underlying action and provides a hierarchy of consistency tests beyond conventional equation of state constraints. In particular, we derived a direct relation between $\mathcal{F}_0$, $H_0$, and $\sigma_8$, showing that the mapping between expansion and growth is controlled by the microphysical flow and this, crucially, elevates $\mathcal{F}_0$ to a central quantity in the discussion of cosmological tensions. The framework developed here could have far reaching implications. It provides a pathway toward systematically reducing the dark energy theory space by imposing increasingly stringent geometric and dynamical constraints. In an upcoming work, we shall be extending this idea further and introducing the misalignment interpretation of cosmological tensions. In this picture, consistency of cosmological data requires not only agreement in individual observables, but alignment in the underlying flow structure which future observations can test through joint reconstructions of $w(z)$ and $c_s^2(z)$.
\\
\\
\section*{Acknowledgements}
We gratefully acknowledge support from Vanderbilt University and the U.S. National Science Foundation. The work of OT is supported in part by the Vanderbilt Discovery Doctoral Fellowship. The work of AG is supported in part by NSF Award PHY-2411502.

\bibliography{apssamp}% Produces the bibliography via BibTeX.

\bibliographystyle{apsrev4-2}
\end{document}